\documentclass{emulateapj}
\begin{document}

\title{The Effect of Starburst Metallicity on Bright X-Ray Binary Formation Pathways}
\author{T. Linden\altaffilmark{1,2}, V. Kalogera\altaffilmark{2}, J. F. Sepinsky\altaffilmark{2,3}, A. Prestwich\altaffilmark{4},  A. Zezas\altaffilmark{4}, J. Gallagher\altaffilmark{5}}
\affil{$^1$ Department of Physics, University of California, Santa Cruz, 1156 High Street, Santa Cruz, CA, 95064, USA}
\affil{$^2$ Department of Physics and Astronomy, Northwestern University, 2145 Sheridan Road, Evanston, IL 60208, USA}
\affil{$^3$ Department of Physics and Electrical Engineering, The University of Scranton, Scranton, PA 18510, USA, USA}
\affil{$^4$ Harvard-Smithsonian Center for Astrophysics, 60 Garden Street, Cambridge, MA 02138, USA}
\affil{$^5$ Department of Astronomy, University of Wisconsin, Madison, WI 53706-1582, USA}

\slugcomment{Submitted to ApJ}
\shortauthors{}

\begin{abstract}
We investigate the characteristics of young ($<$~20~Myr) and bright (L$_X$ $>$ 1 x 10$^{36}$ erg s$^{-1}$) high-mass X-ray binaries (HMXBs) and find the population to be strongly metallicity dependent. We separate the model populations among two distinct formation pathways: (1) systems undergoing active Roche lobe overflow (RLO), and (2) wind accretion systems with donors in the (super)giant (SG) stage, which we find to dominate the HMXB population. We find metallicity to primarily affect the number of systems which move through each formation pathway, rather than the observable parameters of systems which move through each individual pathway. We discuss the most important model parameters affecting the HMXB population at both low and high metallicities. Using these results, we show that (1) the population of ultra-luminous X-ray sources can be consistently described by very bright HMXBs which undergo stable RLO with mild super-Eddington accretion and (2) the HMXB population of the bright starburst galaxy NGC~1569 is likely dominated by one extremely metal-poor starburst cluster.
\end{abstract}

\section{Introduction and Background}

High-mass X-Ray binaries (HMXBs) form an important observational tool in studies of star-forming regions, due to both their relatively high X-Ray luminosity and age-specificity, which make them good indicators of recent star formation \citep{2003MNRAS.339..793G}. In extragalactic environments, often only \emph{bright} HMXBs (here defined as L$_X$ $>$ 1 x 10$^{36}$ erg s$^{-1}$) are detectable. This may significantly affect not only the absolute number of sources, but also the time evolution and observable characteristics of discoverable HMXBs. 

Recent studies have noted an inverse correlation between starburst metallicity and the number of observable HMXBs produced by each starburst. \citet{2004ApJ...609..133M} find the Small Magellanic Cloud (SMC) to be overabundant in HMXBs by a factor of $\sim$50 compared to the Milky Way - though they note that several factors, including (1) incompleteness in Milky Way observation, (2) recent star formation in the SMC, and (3) the large number of Be-HMXBs observed in the SMC, greatly contribute to this result. Using population synthesis techniques to model continuous star formation activity, \citet{2006MNRAS.370.2079D} found a factor of $\sim$3 increase in the total number of HMXBs with L$_X$~$>$~1~x~10$^{32}$~erg~s$^{-1}$ when the metallicity is decreased from solar (Z=Z$_\odot$) top SMC (Z=0.2~Z$_\odot$).

Within this population of bright extragalactic HMXBs exists a subpopulation of ultra-luminous X-ray sources (ULX; defined here as L$_{X}$ $>$ 1~x~10$^{39}$~erg~s$^{-1}$, the Eddington limit for a $\simeq$ 7 M$_\odot$ black hole), the exact nature of which is not well understood \citep{1997ApJ...478..542F, 2000ApJ...535..632M, 2001ApJ...554.1035F, 2004PThPS.155...27M, 2004ApJS..154..519S}. Because some of these systems have isotropic luminosities above the Eddington luminosity for black holes (BHs) thought to be produced via single star evolution, much work has centered on the possibility that ULXs contain an intermediate-mass BH (IMBH) \citep{1999ApJ...519...89C, 2004cbhg.symp...37V, 2008ApJ...688.1235M}. Other arguments point to ULXs as a natural extension of the HMXB population \citep{2003MNRAS.339..793G}. These models may bypass the strict isotropic Eddington limit by invoking photon bubbles which can escape from a ``leaky disk" \citep{2002ApJ...568L..97B}, a non-spherical component of the X-Ray luminosity (beaming) \citep{2001ApJ...552L.109K} or an optically thin accretion disk \citep{1988ApJ...332..646A, 2003ApJ...597..780E}. These stellar mass ULXs may achieve super-Eddington luminosities through either strong wind accretion from supergiant donors \citep{2005MNRAS.356...12S} or thermal timescale Roche-lobe overflow (RLO) \citep{2001ApJ...552L.109K}. 

Observations have found a correlation between ULXs and young starburst activity \citep{2004ApJ...601L.143I}. The most recent observations have suggested that the number of ULX sources peak between 10-20 Myr after star formation has ended \citep{2009ApJ...703..159S}. These observations have also found ULX to be most prevelant in low metallicity environments \citep{2002astro.ph..2488P, 2004ApJ...603..523Z, 2008ApJ...684..282S}. Several models have been posited to explain the metallicity dependence of ULX sources. \citet{2009MNRAS.395L..71M}, and later \citet{2009MNRAS.400..677Z} have employed the results of evolutionary calculations that indicate an enhanced formation of high mass black holes ($\sim$ 30 M$_\odot$) at low metallicities \citep{2006A&A...452..295E,2010arXiv1004.0386B}. The large masses of these BHs provide them with Eddington limits which reach typical ULX luminosities ($\sim$ 10$^{40}$~erg~s$^{-1}$). These massive stellar  black holes are expected to form through direct stellar collapse witn no associated natal kick (e.g., \citet{2001ApJ...554..548F, 2003Sci...300.1119M}). However, \citet{2002ApJ...577..710Z} found a moderate displacement between ULX sources and young stellar clusters in the Antennae galaxies, suggesting that the ULX population received a natal kick, indicating that these systems either did not undergo direct collapse, or that direct collapse events are still associated with a small to moderate natal kick. 

Seeking to bridge the gap between theoretical predictions and observation, in this \emph{paper} we analyze the formation of bright HMXBs in the first 20 Myr after starburst activity at several different metallicities. We show that, while our models can produce bright HMXBs at all metallicities, the number of systems, time evolution, and orbital period distribution are strongly metallicity dependent. In particular, while our research supports the observation that ULX are preferentially formed in low-metallicity environments, we find that this relation is not related to the formation of large ($\sim$~30~M$_\odot$) direct-collapse BHs at low metallicities, but instead to the dynamics of the pathway which allows RLO onto black holes of $\sim$10-15 M$_\odot$ BHs, with mild super-Eddington luminosities. 

In Section~\ref{modeling} we describe our simulation codes and modeling assumptions. In Sections~\ref{sec:results} and \ref{sec:pathways} we analyze our results on the properties of bright HMXBs and discuss their physical origin in the context of their evolutionary formation pathways. In Section~\ref{theorycompare} we discuss the relations between our results and previous work, specifically noting the results for the ULX population in Section~\ref{subsec:ulx}. Further applying our results to observed systems, in Section~\ref{subsec:ngc1569} we discuss the production of HMXBs in the starburst galaxy NGC 1569 in the context of the ages and metallicities of the known super star clusters. Finally, we summarize our conclusions in Section~\ref{sec:conclusion}.

\section{Simulation Code and Models}
\label{modeling}

Employing a sophisticated and recently updated population synthesis code, $StarTrack$ \citep{2002ApJ...572..407B, 2008ApJS..174..223B}, we generate HMXBs from a variety of input parameters using Monte Carlo methods. We note that the parameter space for Monte Carlo population synthesis of HMXB systems is formidable in size: thus a full exploration of the parameter space is not computational feasible. We take default values of many parameters, following the layout set forth in \citet{2008ApJS..174..223B}.

For our initial conditions, we employ a Salpeter (M$^{-2.35}$) initial mass function, setting the minimum primary mass at 4~M$_\odot$ and choosing the secondary mass via a flat mass ratio distribution with a minimum secondary mass of 0.08~M$_\odot$. Following the prescription of \citet{1983ARA&A..21..343A}, we create a distribution of initial binary separations which is flat in the logarithm with an upper limit of 10$^5$ R$_\odot$ and a lower limit such that the primary star initially fills at most half of its Roche Lobe. We employ a thermal distribution for initial eccentricities \citep{1975MNRAS.173..729H}. 

The StarTrack code then models the evolution of each individual star using the stellar tracks modified from those used in the Single Star Evolution (SSE) code \citep{1998MNRAS.298..525P, 2000MNRAS.315..543H}. These models include metallicity dependent evolution parameters which affect the radius, temperature, and wind mass loss rates of each star in our simulation. We also employ a revised model for wind accretion from massive stars (\citet{2010arXiv1004.0386B}, which employs modifications to adapt for the results of \citet{2001A&A...369..574V}). 

The stellar characteristics of each star are modified through their binary interactions. The most important stellar effects come from stable mass transfer through Roche Lobe Overflow (RLO) and unstable common envelope (CE) phases. Stable RLO phases can be treated by comparing the corresponding timescale for mass transfer from angular momentum losses and nuclear evolution of the donor with the donor's thermal timescale. As long as the donor remains in thermal equilibrium the mass transfer rate is then stable \citep[Section 5.1]{2008ApJS..174..223B}. CE phases are treated using the energy formalism \citep{1984ApJ...277..355W}, where the binary is immediately circularized and the amount of energy needed to remove the envelope of the primary star is subtracted from the potential energy of the binary orbit. If this process results in a stable binary system, we continue to track the evolution of the binary system. However, if there is not enough orbital energy in the binary to remove the envelope without creating a merger of the two cores, we immediately end the simulation. For main sequence (MS) donors, the common envelope is invariably assumed to lead to a binary merger. Hertzsprung gap (HG) stars have not yet fully formed clear core-envelope structure and the inspiral process in the CE phase is very likely to lead to a binary merger \citep{2000ARA&A..38..113T}. However, due to the uncertainties in the common envelope input physics, we either assume that all binaries merge in CE initiated by HG star (our default models) or are allowed to survive provided that there is enough orbital energy to eject envelope (alternative model). For further details on the importance of this parameter choice to the formation of binary systems, see \citet{2007ApJ...664..986B}.

We set the maximum neutron star (NS) mass at 2.5~M$_\odot$ and the minimum CO core mass for direct collapse BHs to be 7.6~M$_\odot$ \citep{1999ApJ...522..413F, 2001ApJ...554..548F}. At the time of compact-object formation, we assign a random natal kick drawn from a single Maxwellian distribution with a dispersion velocity of 265~km~s$^{-1}$ \citep{2005MNRAS.360..974H}. We scale down the natal kick assigned to BHs by multiplying the kick by the fraction of material which does not fall back onto the compact object. In situations where BHs form through direct collapse (e.g., where the entirety of the supernova ejecta falls back onto the BH and there is no mass loss other than the BH binding energy), we apply no natal kick in our default models.

We note that a ULX cutoff of L$_{X}$ = 1 x 10$^{39}$ erg s$^{-1}$ sets an Eddington-limited BH mass of $\sim$7 M$_\odot$. However, several models allowing super-Eddington accretion have been postulated \citep{1988ApJ...332..646A, 2001ApJ...552L.109K, 2002ApJ...568L..97B}. Instead of assuming a specific physical model for super-Eddington emission, we simply set the luminosity to be the lesser of (1) the luminosity calculated without regard to the Eddington limit or (2) ten times the Eddington limited luminosity for BH accretors and two times the Eddington limited luminosity for NS accretors. Such super-Eddington luminosities for BHs have been found in simulations of slim-accretion disks \citep{1988ApJ...332..646A, 2002ApJ...568L..97B, 2003ApJ...597..780E, 2006ApJ...651.1049S} and photon bubble instability \citep{2003ApJ...586..384R} These effects may even allow violations of the Eddington limit by a factor as large as 100 \citep{2002ApJ...568L..97B, 2006ApJ...643.1065B} so our super-Eddington model may remain conservative. On the other hand, NS accretors seem to remain within twice the Eddington rate~\citep{1991ApJ...381..101L,1993ApJ...410..328L, 2003ChJAS...3..257G}. This approximation has subsequently been employed in several theoretical models of luminous X-Ray binary formation \citep{2005MNRAS.356..401R, 2008ApJ...683..346F}, and we adopt this assumption throughout. In order to compare our results with {\em Chandra} observations, we apply an estimated correction for the {\em Chandra} energy band as described and justified in Section 9.1 of \citet{2008ApJS..174..223B}. 

We make two important parameter space choices throughout this work. We first choose to evaluate only systems younger than 20 Myr of age. This cutoff corresponds both to the first appearances of electron-capture supernovae (ECS) \citep{1984ApJ...277..791N, 2008ApJ...675..614P} in our simulations as well as the last iron-core collapse SN. In \citet{2009ApJ...699.1573L}, we have found that ECS are highly associated with the production of Be-HMXBs. Regardless of the accuracy of this association, we note that \citet{2005ApJS..161..118M} found the observed number of Be-HMXBs to be greatest between 25-100 Myr after star formation, relegating the vast majority of Be-HMXB activity until after the end of the simulations and the questions considered here. Be-HMXBs have been found to dominate well studied systems, such as the SMC, with transient luminosities exceeding our bright luminosity cutoff \citep{2000A&AS..147...25L, 2008arXiv0812.1226A}. However, current models of non-spherical winds from Be donors are not well refined, and thus the number and duration of bright Be-HMXB activity is a major source of uncertainty for our models. Secondly, we have chosen a luminosity cutoff of L$_{X}$~=~1~x~10$^{36}$~erg~s$^{-1}$, as this is the sensitivity limit of many $Chandra$ observations of young starbursts and HMXB populations.

Another possible mechanism for the production of bright HMXBs are short-lived phases of unstable atmospheric Roche Lobe overflow \citep{1979A&A....71..352S,1983adsx.conf..343S}. Due to the very short timescale of atmospheric RLO (on the order of 10$^5$ yr~\citep{1983adsx.conf..343S} compared to 10$^7$ yr for thermal timescale RLO~\citep{2008ApJS..174..223B}), we do not consider these systems in this work. However, in order to estimate the relative error caused by disregarding these systems, we investigate the percentage of atmospheric RLO events which result in stable RLO. Given a Salpeter initial mass function, the average initial mass of SN progenitors in our simulation is $\sim$24 M$_\odot$. BHs formed form these primary stars will lose approximately 2/3 of their initial mass before CO creation. Since we employ a flat secondary mass distribution, we then expect at least 1/3 of initial secondaries to be less massive than the CO - and thus to undergo a thermal timescale RLO-HMXB period after atmospheric RLO. Thus, we expect the omission of atmospheric RLO to create at most a 20\% underestimate in the population of bright RLO-HMXBs in our simulations. Since most RLO-HMXBs already accrete at or above the Eddington limit, we expect no increase in ULX activity from atmospheric RLO.

To asses the major uncertainties of single and binary evolution, we evaluate changes due a number of important parameters individually. The parameters most likely to effect HMXB production include: CE efficiency, the treatment of HG CEs, natal kick distributions, the existence of natal kicks for direct collapse SN, and violations of the Eddington limit. The exact role of each of these parameters will be examined in Section~\ref{sec:pathways}.

\section{Results for Default Model}
\label{sec:results}

\begin{figure}
		\plotone{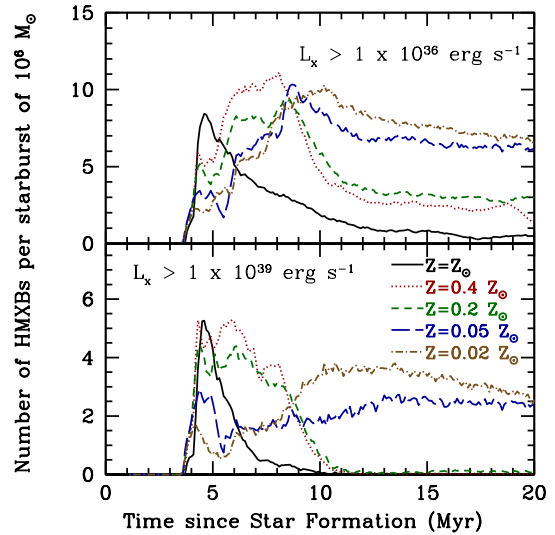}
		\caption{Number of detectable HMXBs at Z=(Z$_\odot$,~0.4Z$_\odot$,~0.2Z$_\odot$,~0.05Z$_\odot$,~0.02Z$_\odot$) with L$_{X}>$~1~x~10$^{36}$~erg~s$^{-1}$ (bright, top) and L$_{X}>$~1~x~$10^{39}$~erg~s$^{-1}$ (ULX, bottom) over the first 20 Myr after a starburst of 10$^6$ M$_\odot$. Numbers are generated as the average of $\sim$122 simulations. (Default evolutionary model, 10x Eddington limit L$_x$ allowed)}
\label{numplot}
\end{figure}

Our models show conclusively that the amplitude (number of systems) and time dependence of HMXB activity are strongly metallicity dependent. In Figure~\ref{numplot}, we show the number of HMXBs with L$_X$~$>$~1~x~10$^{36}$~erg~s$^{-1}$ (bright, top) and L$_X$~$>$~1~x~10$^{39}$~erg~s$^{-1}$ (ULX, bottom) as a function of time for a starburst of 10$^6$~M$_\odot$ at metallicities of Z=(Z$_\odot$,~0.4Z$_\odot$,~0.2Z$_\odot$,~0.05Z$_\odot$,~0.02Z$_\odot$) \footnote{Throughout this paper, we generally refer to systems of Z$_\odot$ and 0.4Z$_\odot$ as high metallicity environments, systems with 0.2Z$_\odot$ as a medium metallicity environment, and systems with 0.05Z$_\odot$ and 0.02Z$_\odot$ as low metallicity environments. This naming scheme is based on the trends we see in the HMXB population as a function of metallicity, but may be at odds with observational conventions.} All computational simulations presented throughout this paper were done for an initial population of 10$^6$ binaries with primary mass greater than 4M$_\odot$, which making assumptions for a binary fraction of 1.0 and a Salpeter IMF, translates to approximately 122 realizations. In all plots we have shown the average of these realizations. Evaluating our population of bright sources over the entire 20 Myr simulation, we find that decreasing metallicity monotonically increases the number of HMXBs by a factor of~$\sim$3.5 between Z=Z$_\odot$ and Z=0.02~Z$_\odot$. At the ULX luminosity cutoff (bottom), the low metallicity sources are even more dominant, with a factor of $\sim$5 change from our lowest to highest metallicity and with a increase of $\sim$2.5 between Z=Z$_\odot$ and Z=0.4~Z$_\odot$. 

The time evolution of our HMXB population is also dependent on the starburst metallicity. During the period between 5-10~Myr after star formation, the metallicity dependence of bright HMXB is non-monotonic, with a peak at Z=0.4~Z$_\odot$ and less than half as many systems at both metallicity extremes. The reasons behind this non-monotonicity are discussed in Section~\ref{sgpathway}. This trend disappears after $\approx$~10~Myr, after which low metallicity systems dominate the bright HMXB population by a factor of 5. We note that while the number of low metallicity systems declines after 20 Myr, they continue to dominate over the high metallicity population, creating an even stronger trend towards low metallicity HMXBs when we continue our simulation to later time periods.

At the ULX cutoff (bottom), we note that after 10~Myr, the high metallicity HMXB population almost entirely disappears, while the low metallicity populations are reaching their peak. In this period, we find HMXBs at Z=0.02~Z$_\odot$ to outnumber Z$_\odot$ HMXBs by a ratio of nearly 1000 to 1 (with considerable statistical uncertainties due to the very small number of high metallicity systems produced). This lies in stark contrast to the approximately 2 to 1 ratio found at our bright luminosity cutoff.

\begin{figure}
		\plotone{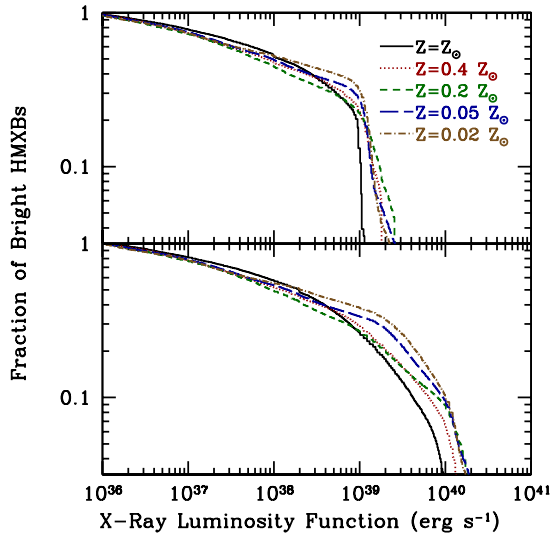}
		\caption{Cumulative X-Ray Luminosity functions for HMXBs at Z=(Z$_\odot$,~0.4Z$_\odot$,~0.2Z$_\odot$,~0.05Z$_\odot$,~0.02Z$_\odot$) when strictly applying the Eddington limit (top) and for models allowing super-Eddington accretion using our default formalism (bottom). Results are taken over the entire 20~Myr simulation and are normalized to the total number of systems with L$_{x}$~$>$~1x10$^{36}$~erg~s$^{-1}$ at each individual metallicity when super-Eddington accretion is enabled. (Default evolutionary model)}
\label{supereddington}
\end{figure}

We note that our determination of HMXB numbers could be greatly influenced by our choice to allow mild violations of the Eddington limit. In Figure~\ref{supereddington} we show the X-Ray luminosity function (XLF) at each metallicity, given both strict adherence to the Eddington cutoff (top) as well as our super-Eddington formalism (bottom). We normalize the number of HMXB at each metallicity to the number above our bright cutoff in our default model. We note that the XLFs in our study are consistent with power-laws up to a luminosity of L$_{X}=$~1~x~$10^{39}$~erg~s$^{-1}$ and are not metallicity dependent, consistent with observations of local galaxies. We caution, however, that our modeled XLF should not be considered as a model for the global HMXB population in galaxies with young or continuous star formation, for several reasons: (1) we consider only HMXBs formed within the first 20 Myr after star formation, while all observable populations contain both young and medium age HMXBs, (2) we plot only the orbit averaged luminosity due to spherical winds, while the majority of observed extragalactic systems are found during burst activity, and (3) observations of the extragalactic XLF \citep[see e.g.][]{2004NuPhS.132..369G} are highly dependent on the assumed SFR in each galaxy, and often have statistical uncertainties due to small number statistics. Given these caveats, careful analysis of the XLF shape and of comparisons to observational XLFs are beyond the scope of this present study, and thus we do not expand further. Such an analysis would require developing models which target specific known galaxies, using their star formation history as model input. Here we instead focus on the brightest HMXBs in starburst environments and how metallicity influences their formation pathways. Finally, we note that the simple power-law explanation underscores the fact that our resultant population should not be highly sensitive to our choice of a luminosity cutoff at L$_{X}=$~1~x~$10^{36}$~erg~s$^{-1}$ - a result which will become clearer as we analyze the pathways through which HMXBs form.

\begin{figure}
		\plotone{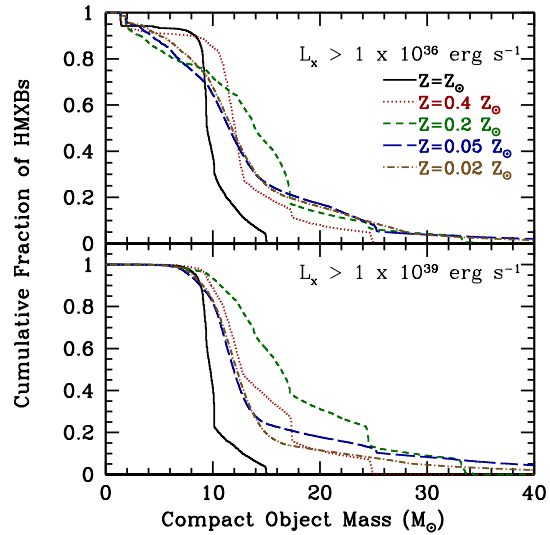}
		\caption{Cumulative distribution of compact object masses for sources at Z=(Z$_\odot$,~0.4Z$_\odot$,~0.2Z$_\odot$,~0.05Z$_\odot$,~0.02Z$_\odot$) for systems with $L_X>10^{36}$\,erg\,s$^{-1}$ (top) and $L_X>10^{39}$\,erg\,s$^{-1}$ (bottom). All results are normalized to the total number of systems throughout each simulation, which here corresponds to 122 realizations of 10$^6$ M$_\odot$ starbursts with an average HMXB number (over 20 Myr) of 5.73 at Z=0.02Z$_\odot$, 5.25 at Z=0.05Z$_\odot$, 3.81 at Z=0.2Z$_\odot$, 3.79 at Z=0.04Z$_\odot$, and 1.71 at Z=Z$_\odot$. The number of systems with compact object masses greater than 40~M$_\odot$ is less than 5~\% at all metallicity and luminosity cuts. We note that the number of systems at ULX cutoff and Z=Z$_\odot$, is very small, and this result is statistically uncertain. (Default evolutionary model, 10x Eddington L$_x$ allowed)}
\label{comasses}
\end{figure}

\begin{figure}
		\plotone{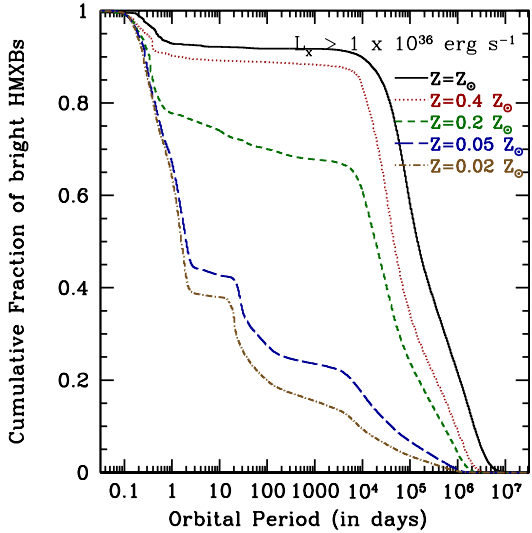}
		\caption{Cumulative distribution of HMXB periods for sources at  Z=(Z$_\odot$,~0.4Z$_\odot$,~0.2Z$_\odot$,~0.05Z$_\odot$,~0.02Z$_\odot$) for systems with $L_X>10^{36}$\,erg\,s$^{-1}$. (Default evolutionary model, 10x Eddington L$_x$ allowed)}
\label{periodplot}
\end{figure}

At all metallicities, we find that our choice to allow super-Eddington accretion increases the total number of systems above our bright luminosity cutoff by less than 2\%. However, Figure~\ref{supereddington} also shows that allowing super-Eddington accretion greatly affects the number of systems above our ULX cutoff. In the Eddington limited case (top), we note a sharp cutoff in the XLF at L$_{x}$~$>$~2~x~10$^{39}$~erg~s$^{-1}$, with nearly 30\% of HMXB having luminosities 1~x~10$^{39}$~erg~s$^{-1}$~$<$~L$_{x}$~$<$~2~x~10$^{39}$~erg~s$^{-1}$. The proximity of these luminosities to our ULX cutoff is not surprising, as we have defined the ULX cutoff to be near the Eddington luminosity for H accretion from a 7~M$_\odot$ BH. However we note that no such cutoff is observed by \citet{2004NuPhS.132..369G}, which is at odds with simulations finding such a large fraction of bright HMXBs radiating at the Eddington limit. When we allow super-Eddington accretion (bottom), we note that between 20-30\% of our bright HMXB population now has a luminosity greater than the L$_{x}$~$>$~2~x~10$^{39}$~erg~s$^{-1}$ cutoff, with a new power law of approximately 0.75. This implies that many of our Eddington-limited systems have mass transfer rates capable of super-Eddington accretion if the accretors are able to accept this additional mass. 

In order to further analyze the observable parameter space of our bright HMXBs, in Figure~\ref{comasses} we plot the masses of the compact objects (COs) in our models, noting several important results. First, we see very few NSs (systems of 1-2~M$_{\odot}$) at our bright luminosity cutoff ($<$~5\%) and no NSs our our ULX cutoff. More importantly, the ULX population does \emph{not} have significantly more massive accretors than the bright HMXB population. At low metallicity, the vast majority of ULX accretors lie between 10-15 M$_\odot$. We note that although simulations show that low metallicities tend to favor large single-star BH masses \citep{2006ApJ...650..303B, 2009arXiv0904.2784B}, the BH masses we observe in low metallicity environments does not approach the maximum possible masses from single star models. Instead we find that the largest BH masses correspond to moderate metallicities (Z=0.2~Z$_\odot$). This absence of massive ULX accretors lies in direct contrast to earlier discussions of reasons for increased ULX production in low-metallicity environments \citep{2009MNRAS.395L..71M, 2009arXiv0909.1017Z}. We will further analyze this issue in Section~\ref{subsec:ulx}.

These results demonstrate the substantial variation between bright and ULX HMXBs, as well as high and low metallicity binaries. The metallicity dependent time evolution at both luminosity cuts in our simulation indicates that we are likely dealing with an HMXB population which is composed of several distinct sub-groups. In order to analyze the trends which affect our bright HMXB population, we will now examine the evolution pathways which are responsible for both bright and ULX HMXB activity. 

\section{Pathways and Model Variation}
\label{sec:pathways}

\begin{figure*}
		\plotone{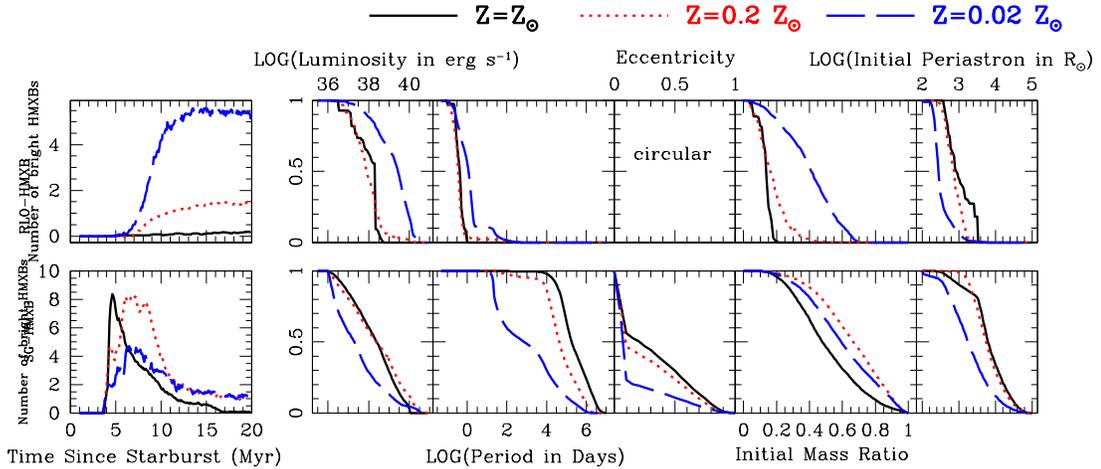}
		\caption{Categorization of HMXBs which shows the (Column 1) time dependent HMXB number (2) cumulative X-Ray luminosity function (3) cumulative orbital period distribution (days), (4) cumulative eccentricity distribution (5) cumulative initial mass ratio distribution (secondary mass/primary mass), (6) cumulative initial periastron separation (in R$_\odot$) for systems from (Row 1) RLO-HMXBs (2) SG-HMXBs. Each plot except for the first is normalized to the number of systems moving through each pathway at each given metallicity. For readability, data are shown only at reference metallicities of Z=Z$_\odot$, Z=0.2~Z$_\odot$, Z=0.02~Z$_\odot$. (Default evolutionary model, 10x Eddington L$_x$ allowed)}
		\label{pathwayplot}
\end{figure*}

We noted several interesting trends in Section~\ref{sec:results} which indicate that our population of HMXBs may not have uniform characteristics. However, the smoking-gun demonstrating the metallicity dependence of the bright HMXB population is shown in Figure~\ref{periodplot}, where we plot the cumulative period distribution of HMXBs at each of our test metallicities. We note that there is a stark contrast between the orbital periods of the HMXB population at Z=0.02~Z$_\odot$ and Z=0.05~Z$_\odot$ and the population at Z=0.4~Z$_\odot$ and Z=Z$_\odot$, with Z=0.2~Z$_\odot$ acting as an intermediate case. For low metallicity systems, the vast majority ($>$60\%) have periods of less than 3 days. while 90\% of high metallicity systems have a period of greater than 10,000 days. At the Z=0.2~Z$_\odot$ metallicity, approximately 20\% of the systems have periods of less than 1 day, with the remaining 70\% of systems having periods of more than 10,000 days. We note that at high metallicities, there are almost no intermediate systems of periods between 1-10,000 days, while at low metallicity, this accounts for approximately 20\% of bright HMXB. 

In order to isolate the effect of metallicity on our HMXB population, we analyze our model populations of bright HMXBs and identify six distinct binary evolution pathways: (1) systems undergoing stable mass transfer via Roche lobe overflow (RLO-HMXB), (2) systems with wind accretion from a MS donor that has moved through a CE phase, (3) systems with wind accretion from a MS donor that has \emph{not} moved through a CE phase, (4) systems with wind accretion from a (super)giant donor (SG-HMXB), (5) systems with wind accretion from a He-rich donor that has moved through a CE phase, and (6) systems with wind accretion from a He-rich donor that has \emph{not} moved through a CE phase. Given these conditions, each bright HMXB falls into exactly one of these catagories at each moment in time, although HMXBs can move between different populations over time. We note that these pathways are very similar to those first described by \citet{2000A&A...358..462V} in their study of the important HMXB contributions to the galactic X-Ray luminosity function (XLF), with agreement on the dominant pathways in the high luminosity regime. Throughout the rest of this section, we will restrict ourselves to examining metallicities of Z=Z$_\odot$, Z=0.2~Z$_\odot$ and Z=0.02~Z$_\odot$ for simplicity, as we find these metallicities to adequately demonstrate the main effects and dependencies of our models. 



Due to the high luminosity cutoff of 1~x~10$^{36}$~erg~s$^{-1}$, our simulations do not find any wind accretion systems with main sequence or Helium rich donors. We note that the spherical winds of MS and He-rich stars are not generally strong enough to promote such extreme HMXB luminosities. However non-spherical winds (e.g., due to rotation) may create bright systems which our models would not account for. Specifically, observations point towards the existence of a large population of transient HMXBs with Be donors \citep{2000A&AS..147...25L, 2008arXiv0812.1226A} that can temporarily exceed our luminosity cutoff. However, as noted in Section~\ref{modeling}, we have chosen here to study only systems younger than 20~Myr, where we believe the observed sample to contain few Be-HMXBs. 

Thus, we find our bright HMXB population to be dominated by systems moving through either the RLO-HMXB or the SG-HMXB pathways. In order to better understand the intial and final conditions which create systems of each class, in Figure~\ref{pathwayplot} we show various observable parameters and initial conditions for systems moving through each of our two dominant pathways. We immediately note two important results: (1) low-metallicity HMXBs are dominated by systems moving through the RLO-HMXB pathway, while high-metallicity HMXBs are dominated by systems moving through the SG-HMXB pathway; (2) to a rough approximation, metallicity does not greatly affect the observable properties of systems moving through a given pathway, instead it affects the number of HMXBs in that pathway. In order to better understand the physics which controls each HMXB pathway, we will individually examine each channel and explain the physical properties that create the observable HMXB population from the typical progenitor parameter space for each channel.

\subsection{RLO-HMXB Pathway}
\label{rlopathway}

Roche lobe overflow provides an important method for creating bright HMXBs (with typical luminosities above L$_X$~$>$~1~x~10$^{38}$~erg~s$^{-1}$) due to its effectiveness in transferring matter onto the compact object accretor. In order to obtain stable mass transfer through RLO, these HMXBs must have both short orbital periods (often $<$ 1 day, Figure~\ref{pathwayplot}, row 1, column 3), and a mass ratio that is near unity during the HMXB phase. Due to the existence of stable mass transfer, these systems are assumed to be circularized in our models (for a discussion of this assumption, see \citet{2007ApJ...667.1170S}).

In order to create HMXBs with small orbital periods, all RLO-HMXBs evolve through a CE phase prior to the supernova of the primary star. CE phases are necessary for RLO-HMXB progenitors because the orbital separation necessary to survive a strong SN natal kick ($\sim$~10~R$_\odot$) is substantially less than the orbital separation necessary to prevent the binary merger of unevolved massive stars ($>$~100~R$_\odot$ at all metallicities). Furthermore, a very tight binary system (typically $<$~5~R$_\odot$) is necessary to allow eventual RLO of the donor star. Because the creation of CEs requires a large mass ratio between the primary and secondary stars (Figure~\ref{pathwayplot}, row 1, column 5) the majority of RLO-HMXB systems ($>$~90\% at all metallicities) contain an unevolved donor accreting onto the compact object. 

\begin{figure}
		\plotone{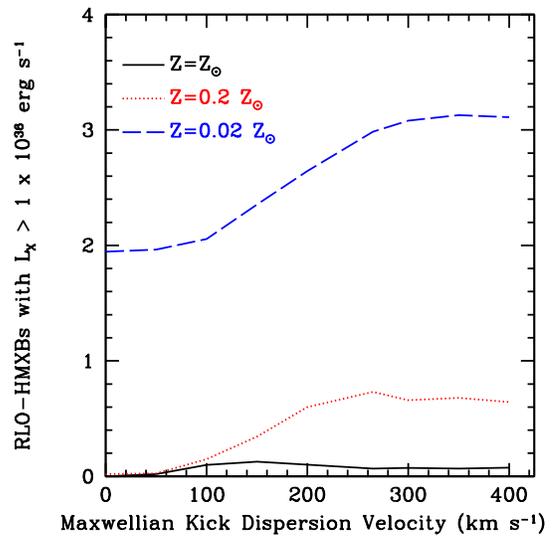}
		\caption{Average number of RLO-HMXBs with $L_X>10^{36}$\,erg\,s$^{-1}$ over the first 20 Myr after a starburst of 10$^6$ M$_\odot$. for Maxwellian natal kick distributions with different dispersion velocities (Natal kick velocity varies from default model, 10x Eddington L$_x$ allowed)}
\label{mtnumvkick}
\end{figure}

\begin{figure}
		\plotone{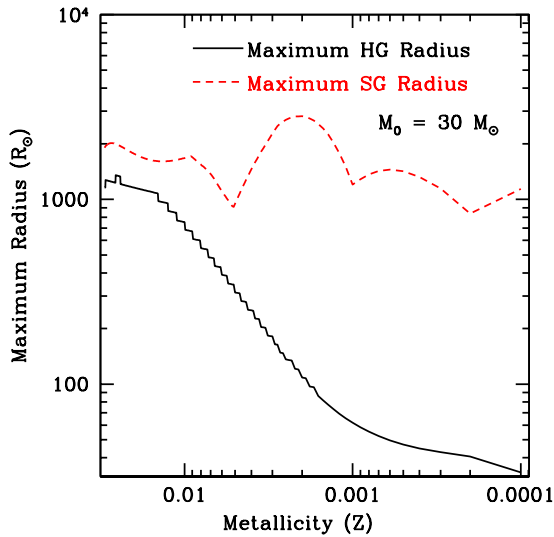}
		\caption{Maximum radius of stars during the HG (solid black) and (super)giant (dashed red) stages for stars of varying metallicity, as determined by the SSE code \citep{2000MNRAS.315..543H}}
\label{sseplot30}
\end{figure}

The dominance of the CE driven pathway explains the time evolution of RLO-HMXBs. We do not begin to see bright systems until after 6~Myr, the lifetime of the heaviest stars which are able to enter the (super)giant phase (and thus undergo survivable CE evolution according to \citep{2000ARA&A..38..113T}). The number of bright RLO-HMXBs continues to increase as the number of primary supernovae increase, since no evolution of the donor star is required. This upward trend ends at $\approx$~13~Myr, corresponding to the latest forming supernovae in our simulation. The number of RLO-HMXBs exponentially decays after 20~Myr as the donors begin to move off of the Main Sequence. This characteristic evolution is observed at all metallicities. 

However, the lack of donor evolution introduces a problem for the creation of RLO-HMXBs. The end product of CE evolution should be a He-rich primary alongside a MS secondary. The He-rich primary is always at least as massive as the eventual CO. However, the system must not be in RLO at the end of the CE phase, or the system would be considered a merger via the energy formalism. Since the evolution of the donor star is minimal, another mechanism must be responsible for moving the He-rich/MS system into a CO/MS system in RLO. This extra force is provided by the natal kick provided to the system at CO formation. While the natal kick will likely add energy to the binary system, it's more important contribution is to take the circularized post-CE binary, and move it into a highly eccentric orbit. Even a more distantly separated system is thus likely to be in RLO during periastron - and the frictional forces of RLO are then assumed to circularize the orbit and create a tightly bound system in perpetual RLO \citep[see][]{2007ApJ...667.1170S}.  In Figure~\ref{mtnumvkick}, we show the number of RLO-HMXBs for different Maxwellian dispersion velocities for the natal kicks imparted to our CO population. We note that the number of systems increases significantly when larger natal kicks are employed. At high metallicities, almost no RLO-HMXBs exist when natal kicks are disabled. We note a similar correlation between natal kicks and the number of RLO X-Ray Binaries has been previously theorized in the case of low mass X-Ray binaries \citep{1998ApJ...493..351K}. 

At our Z=0.02~Z$_\odot$ metallicity, we note a persistent class of HMXBs which exist even when natal kicks are disabled. This residual, which contains almost 50\% of the total number of systems, is composed of compact objects resulting from the direct collapse of particularly massive BH progenitors into BHs of 10-15~M$_\odot$. Interestingly, we find this class of systems to exist only when the wind prescription of \citet{2001A&A...369..574V} is employed. Compared to the formulas of \citep{2000MNRAS.315..543H}, the Vink formulation predicts significantly lower wind mass loss in the low metallicity regime. This allows significantly less massive ZAMS stars to end their lives as fallback BHs, compared to systems in the high metallicity regime. These ZAMS stars can be small enough to move through a SG phase and form a tight binary orbit with a MS donor. This region of parameter space is not available when either stronger wind models are used or when the stellar metallicity is high, as the minimum ZAMS mass necessary to create a fallback BH is then larger than the maximum ZAMS mass necessary to move through a stable (super)giant phase.  Upon the formation of the BH, the small mass loss from the binding energy of the progenitor star creates an eccentricity that brings the system into RLOF and creates a bright HMXBs.

The dominance of the CE based pathway for RLO-HMXB formation explains the inverse correlation between RLO-HMXB number and metallicity. In order for a CE to form, the Roche Lobe of the primary star must lie in between the maximum stellar radius during the HG and the maximum stellar radius during the SG phase. If the Roche lobe is smaller than the maximum HG radius, the system will undergo a CE during the HG phase which we assume leads to binary merger. If the Roche lobe is larger than the maximum SG radius, then the system will never enter a CE phase, and thus will not form a RLO-HMXB. The survival of high mass binaries through the common envelope phase at low metallicity was explained in detail by \citet{2010arXiv1004.0386B} in the context of double compact object formation.

\begin{figure}
		\plotone{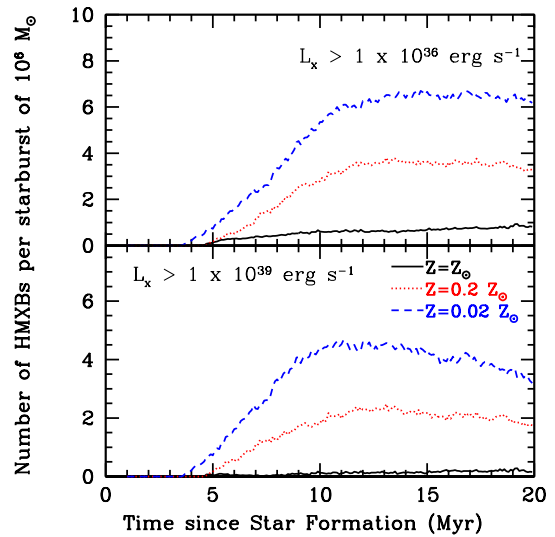}
		\caption{Average number of HMXBs at Z=(Z$_\odot$, 0.2Z$_\odot$, 0.02Z$_\odot$) for L$_{X}>$~1~x~10$^{36}$~erg~s$^{-1}$ (bright, top) and L$_{X}>$~1~x~$10^{39}$~erg~s$^{-1}$ (ULX, bottom) over the first 20 Myr after a starburst of 10$^6$ M$_\odot$ for systems going through RLO-HMXB phases with Hertzsprung Gap common envelope evolution enabled using the energy formalism. (HG CEs allowed, 10x Eddington L$_x$ allowed)}
\label{hgceplot}
\end{figure}

\begin{figure}
		\plotone{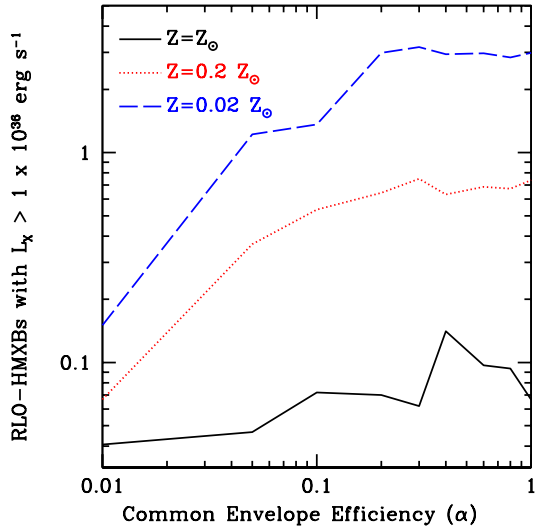}
		\caption{Average number of RLO-HMXBs with $L_X>10^{36}$\,erg\,s$^{-1}$ over the first 20 Myr after a starburst of 10$^6$ M$_\odot$. for different values of the Common Envelope efficiency}
\label{ceplot}
\end{figure}

In Figure~\ref{sseplot30}, we show an illustrative plot from the Single Star Evolution code \citep{2000MNRAS.315..543H} of both the maximum HG radius (ktype = 2 in SSE) and the maximum SG radius (ktype = 4) as a function of metallicity for star with a Zero Age Main Sequence mass of 30~M$_\odot$, which is typical for the primary star in RLO-HMXB progenitors. We note that while the maximum SG radius remains large even at low metallicities, the maximum radius during the HG phase drops precipitously with decreasing metallicity. These curves place strict upper and lower limits on the periastron Roche lobe separation of the binary system. We note that our initial parameter space of orbital separations is set to be flat in the logarithm \citep{1983ARA&A..21..343A}. Since the necessary binary separation is simply a multiplicative constant times the Roche Lobe radius (depending on the mass ratio of the system), the parameter space of allowable stable SG CEs is much wider at low metallicity than at near solar metallicity. While not all of these CEs are survivable via the energy formalism, the trend still creates a large preference towards low metallicity RLO-HMXBs. 

Since HG CEs eliminate a large fraction of the parameter space for RLO-HMXB formation at high metallicities, it is worthwhile to determine the effect of removing this restriction. In Figure~\ref{hgceplot}, we investigate the importance of this assumption by showing the number of bright (top) and ULX (bottom) RLO-HMXBs as a function of time for a simulation where we allow CEs in the HG phase to survive if they are energetically stable. We note two results:  (1) the number of RLO-HMXB systems increases at all metallicities, as all systems with a HG primary were previously assumed to merge, while the new analysis instead allows some to survive and (2) the first RLO-HMXB systems form at much earlier time periods than in our default simulation, as primaries too massive to move through a (super)giant phase can now go through a stable common envelope phase before the 6 Myr genesis of (super)giant systems.

The effects of allowing HG CEs can be drastic, especially at high metallicities, where so few RLO-HMXB progenitors survive the CE during the SG phase. At Z=Z$_\odot$, we see a factor of 6 increase in the number of RLO-HMXBs at our bright luminosity cutoff, and we now see a non-negligible number of HMXBs at the ULX cutoff. At Z=0.2~Z$_\odot$, we see an factor of 3 increase at our bright cutoff, and a factor of 30 increase at our ULX cutoff. At Z=0.02~Z$_\odot$ we see a 20\% increase at our bright cutoff and a 40\% increase at our ULX cutoff.  While \citet{2000ARA&A..38..113T} make a convincing case that HG-CE events should lead to mergers, the sensitive dependence of bright and especially ULX HMXB number on the onset of survivable CE phases demands further studies into this parameter. 

Another large theoretical uncertainty in the physics of CE phases involves their efficiency in transferring the potential energy from the binary orbital separation into ejecting the envelope of the donor star. In Figure~\ref{ceplot}, we show the number of bright RLO-HMXBs as a function of $\alpha$ (defined in  \citet{1984ApJ...277..355W}). We note that RLO-HMXB number is highly conserved at all metallicities for all choices of $\alpha$~$\ge$~0.2. We find that while changing the value of $\alpha$ may determine whether \emph{specific} RLO-HMXB progenitors survive the CE phase, the parameter space of RLO-HMXB progenitors is filled with ``near miss" systems (i.e. systems slightly too close to survive the CE phase, or systems slightly too far apart to enter RLO after the CE phase) such that the number of systems which become RLO-HMXB remains relatively constant with changes in the CE efficiency. In this work we do not consider CE efficiencies greater than unity, as no realistic physical mechanism has been proposed to allow for energy loss greatly in excess of the energy available from the gravitational potential. We note that these Common envelope efficiencies are in agreement with those predicted by~\citet{2000ARA&A..38..113T}.

\subsection{SG-HMXB Pathway}
\label{sgpathway}

From Figure~\ref{pathwayplot} it is clear that SG-HMXBs occupy locations in both observable and initial parameter space that are distinct from the RLO-HMXB population. While the systems have a luminosity function which is similar to RLO-HMXBs, these systems have typical orbital periods greater than 1000 days, and occasionally longer than 10$^5$ days at high metallicity. While a significant fraction ($\approx$ 50\%) of the systems have circularized due to tidal interactions, there is also a significant fraction of systems with moderate eccentricities. 

This observable parameter space is reasonable considering the nature of the SG donors. Since wind accretion is much less efficient than RLO at moving mass onto a CO, only the strongest donor winds can provide the necessary mass loss rates to achieve bright HMXB activity. While highly non-spherical winds may create significant wind accretion with lower mass loss rates, the necessary spherical wind strengths are provided only by (super)giant systems moving though the SG phase. However, as shown in Figure~\ref{sseplot30}, (super)giant systems massive enough to evolve before 10~Myr have extremely large radii ($>$~1000~R$_\odot$), which impose a strict lower bound on the orbital period of the SG-HMXB.

The necessary orbital separations for SG-HMXBs require a specific evolutionary chanel for their progenitors. Unlike RLO-HMXBs, these systems do not move through a CE before the SN of the primary star. A CE driven pathway would create much smaller orbital separations ($\approx$~10~R$_\odot$) than the necessary separation for the SG-HMXB, causing the system to merge before the SG donor has formed. Instead, most SG-HMXB progenitors begin with a large periastron separation ($>$~1000~R$_\odot$ at Z=0.02~Z$_\odot$ and $>$~5000~R$_\odot$ at Z=Z$_\odot$) and undergo no significant binary interactions before the primary SN. A possible exception to this rule is the case of light mass transfer for binaries with mass ratios close to unity. 

\begin{figure}
		\plotone{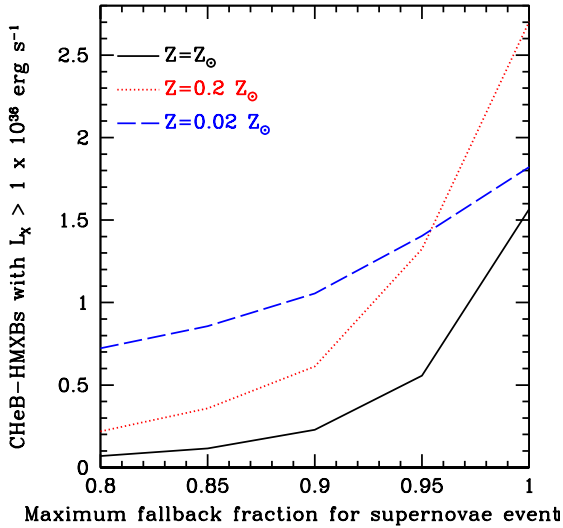}
		\caption{Average number of SG-HMXBs with $L_X>10^{36}$\,erg\,s$^{-1}$ over the first 20 Myr after a starburst of 10$^6$ M$_\odot$ for different values of the maximum fallback fraction for black hole supernovae. (Maximum fallback fraction varies from default model, 10x Eddington L$_x$ allowed)}
\label{chebdc}
\end{figure}

Since the eventual system must have a significant orbital separation at the time of CO formation of the primary, if CO formation were accompanied by an energetic natal kick (with a velocity $\gtrsim$~50~km~s$^{-1}$), the vast majority of SG-HMXB progenitors would be disrupted by the SN. For young systems, the only method for avoiding binary disruption due to natal kicks is to create the CO through direct collapse into a massive BH, which our simulations assume to impart no natal kick to the system. We find that this avenue produces the majority of our SG-HMXBs. 

In Figure~\ref{chebdc}, we show the number of bright SG-HMXBs as a function of the maximum fallback fraction for BHs. Since BHs in our simulations receive kicks scaled down by (1-f$_{fallback})$, this effectively enforces the assignment of a non-zero kick to all BH events. This shows not only that SG-HMXBs form almost exclusively from direct collapse BHs, but that the survival of these systems is radically changed if even small kicks are applied. This restriction is most evidence at high metallicities, where the initial binary separation is largest. We find that adding a 20\% kick to direct collapse events only kills about 60\% of systems at Z=0.02~Z$_\odot$, but destroys 96\% of systems at solar metallicity. Because of this effect, we note that these systems should be found close to the star formation region, since there is no natal kick to impart a large spatial velocity to the binary.

We note that there only exists a short peak ($<$5~Myr) of bright SG-HMXBs. This occurs for three reasons: (1) each SG-HMXB exists only for a short time, because the donor star moves through the SG phase very rapidly ($<$ 1 Myr), (2) nearly all COs are formed before 6 Myr, while direct collapse events are still possible. Given the flat initial mass ratio distribution, we expect most secondary stars to have similar masses to the direct collapse primary, and thus move through the SG stage shortly thereafter, (3) for very small donors, the constraints on the radial separation from the donor star will create systems too dim to be detected above our luminosity cutoff. 

The metallicity dependence of our SG-HMXB population is quite different than that of the RLO-HMXB population. The SG-HMXB population has many systems which fall below our minimum luminosity cut for bright HMXBs. Thus, metallicity-dependent changes in HMXB number occur primarily due to changes in the luminosity of SG-HMXBs. The number of bright systems peaks at our intermediate metallicity of Z=0.2~Z$_\odot$. This is due to two competing effects: (1) stronger winds for high-metallicity donors, and (2) smaller HG radii for low-metallicity donors, allowing SG-HMXBs to form in closer orbits.  

Finally, we note one important parameter space effect which may greatly affect the statistics of our population. Here we have shown only the orbit-averaged luminosities of our SG-HMXB population - while we know that the bright and ULX HMXB population usually exhibits some luminosity variability. From Figure~\ref{pathwayplot}, we note that between 40\%-80\% (for Z=Z$_\odot$ and Z=0.02Z$_\odot$ respectively) of our SG-HMXB population has eccentricities smaller than 0.05, which corresponds to a luminosity variation of less than 22\% from periastron to apastron. At higher resolution, we note that most of these eccentricities are, in fact, almost identically zero, due primarily to tidal effects. However, our remaining SG-HMXB population has a nearly flat eccentricity distribution, which creates a subpopulation of SG-HMXBs with potentially sizable luminosity variations (at least a factor of 50 change for the 10\% of solar metallicity HMXB with eccentricities greater than 0.75). 

There are two effects which limit the impact of HMXB variability on our results - especially in the ULX regime. First, in Figure~\ref{supereddington}, we note that the XLF of our source population is extremely hard. Since HMXBs will only be near periastron for a small fraction of their lifetimes, we would need a large underlying population of HMXBs just under our ULX cut in order for variability to cause these systems to make substantial transient contributions to our ULX population. Secondly, in the same Figure, we note that the creation of our brightest HMXBs is controlled primarily by our allowance of super-Eddington accretion. Thus periastron orbits in our brightest systems will not create significantly brighter HMXBs than we already see - as this luminosity will be capped by our luminosity limit of ten times the Eddington luminosity. However, we note that this variability may have an effect at much lower luminosities, and it may create some variability in the number of observed SG-HMXB sources.

\section{Discussion}
\label{theorycompare}

Our simulations yield a number of key results: (1) the young, extragalactic HMXB population is dominated by systems moving through two specific pathways, RLO of main sequence donors onto reasonably sized BHs, and wind accretion from (super)giant donors onto massive BHs formed via direct collapse (consistent with \citet{2000A&A...358..462V}); (2) while metallicity does not greatly affect the characteristics of HMXBs moving through either pathway, it affects the relative strength (in number of systems) of each pathway, creating a total HMXB population which looks very different at different metallicities; (3) with mild allowances for super-Eddington accretion, the majority of the ULX population can be explained as an extension of the stellar HMXB population (consistent with both observations \citep{2004ApJ...601L.143I, 2002astro.ph..2488P, 2004ApJ...603..523Z, 2008ApJ...684..282S} and theoretical models \citep{2008ApJ...688.1235M, 2009MNRAS.395L..71M, 2009MNRAS.400..677Z}); (4) as opposed to previous models, we find this phenomenon to be connected to the binary evolution of HMXB progenitors, rather than to the fact that low metallicity stars produce more massive BHs.

\subsection{ULX Population}
\label{subsec:ulx}
Re-examining our earlier analysis, we note several observable results in our description of the ULX population due to stellar HMXBs. First, in Figure~\ref{numplot}, we see that the 0.02~Z$_\odot$ environment favors ULX-HMXB formation by approximately a factor of 5 over solar metallicity environments, with a variation of approximately 2 between 0.02~Z$_\odot$ and 0.2~Z$_\odot$. However, this relation is highly time dependent, and higher metallicity environments are favored for very young star forming regions ($<$10~Myr). Since the high metallicity population is dominated by SG-HMXBs, while the low metallicity environment favors RLO-HMXBs, in Figure~\ref{pathwayplot}, we note that high metallicity ULXs should have substantially longer orbital periods than low metallicity systems (see Figure~\ref{periodplot}). 

There are several reasons to believe that our simulated ULX population robustly describes the observed population. In Figure~\ref{supereddington} (top), we note a sudden cutoff in HMXB luminosity at the Eddington limit. This dropoff includes nearly 30\% of our sources with $L_X>10^{36}$\,erg\,s$^{-1}$, and is not observationally detected. Instead, our use of mild super-Eddington accretion (bottom), allows a much smoother decline, which is maintained up to luminosities of 10$^{40}$ erg~s$^{-1}$, consistent with observations of nearby galaxies \citep{2004NuPhS.132..369G}. Secondly, using this method, we can replicate a large percentage of the ULX population, with a reasonable percentage of sources as bright as $L_X~\approx~10^{40}$\,erg\,s$^{-1}$. 

Recent observations have found an inverse correlation between cluster metallicity and ULX formation rates \citep{2007IAUS..238..235S}. It has recently been posited \citep{2009MNRAS.395L..71M} that the reason for this trend is that low metallicity stars lose less mass to stellar winds, and are thus able to produce significantly more massive COs than high metallicity stars. On its face, this hypothesis is reasonable, since the primary restriction on forming ULX systems from the stellar population comes from the Eddington limit, and the Eddington limit increases linearly with CO mass. Thus massive BHs ($\approx$~30~M$_\odot$) would require only mild violations of the Eddington limited luminosity in order to explain the majority of the ULX population. 

While our own modeling codes agree with the correlation between low metallicity and large BH mass \citep[see][]{2006ApJ...650..303B}, there are several problems in correlating this result to explain the ULX population, as done in \citet{2008ApJ...688.1235M, 2009MNRAS.395L..71M, 2009MNRAS.400..677Z}. It is important to note that ULX-HMXBs form from only a very small percentage of the initial binary population, and are thus likely to have properties quite different than the average binary (or average HMXB for that matter). While it is very easy to form large BHs in low metallicity environments, it is actually quite difficult to find them in HMXBs with ULX luminosities for two reasons: (1) the SG-HMXB pathway is unable to create ULX HMXBs because the stellar wind from the low-metallicity donor stars are too weak to achieve these luminosities, and (2) it is difficult to create massive BHs through CE phases, since this tends to strip a large percentage of the primary envelope. 

\subsection{NGC 1569}
\label{subsec:ngc1569}
A second interesting result from our models pertains to the HMXB population of NGC~1569, which is thought to be dominated by clusters less than 20~Myr old \citep{1982ApJS...49...53H, 1988A&A...198..109I, 1991ApJ...370..144W}. Using the $Chandra$ X-Ray telescope, \citet{2002ApJ...574..663M} found 14 X-Ray point sources in NGC 1569, 12 of which are probable X-ray binaries. \citet{2004MNRAS.348L..28K} have found that these X-Ray binaries are found preferentially close to the largest starburst clusters in NGC 1569.

\citet{1985AJ.....90.1163A} found two massive starburst clusters, aptly named Clusters A and B. Both clusters are thought to have ages of less than 20~Myr. We note that \citet{1997ApJ...479L..27D}, and later \citet{2001AJ....122..815O}, used WFPC2 observations to conclude that Cluster A is actually composed of two separate clusters of slightly different ages, a result that is not taken into account here. Later studies by \citet{2000AJ....120.2383H} and \citet{2004MNRAS.347...17A} found 48 and 169 clusters respectively. However, the majority of these clusters have masses several orders of magnitude smaller than Clusters A or B. 

According to the estimates of \citet{2004MNRAS.347...17A}, both Clusters A and B have ages of $~\sim$12 Myr, and masses of 1.64~x~10$^{6}$~M$_\odot$ and 5.65~x~10$^{5}$~M$_\odot$ respectively - though recent studies have shown the Cluster B mass to perhaps be as much as 10\% higher \citep{2008MNRAS.383..263L}. Thus, they would naively be expected to both contribute a significant fraction of the bright HMXB population. However, our results cast an interesting light on the relationship of Clusters A and B to the resultant HMXB population. \citet{2004MNRAS.347...17A} find Cluster A to have a metallicity of Z=0.02~Z$_\odot$, while Cluster B has a much higher metallicity of Z=0.4 Z$_\odot$. From Figure~\ref{numplot}, at a luminosity cutoff of 1 x 10$^{36}$ erg s$^{-1}$ (top), the Cluster A population is preferred by approximately a factor of 3. When accounting for the mass ratio of the two clusters, which is approximately 3 to 1, we find that the HMXB contribution from Cluster A is about 9 times as large as that of Cluster B, meaning that it will almost entirely dominate the HMXB population of NGC 1569. We note that this result would be much different, if Clusters A and B were instead only 8 Myr of age, and the significant SG-HMXB population of Cluster B was still non-negligible.

\section{Conclusion}
\label{sec:conclusion}
We have studied the young, extragalactic, HMXB population and determined that it is composed of systems moving through two distinct binary pathways: mass transfer systems main sequence donors and wind accretion systems containing (super)giant donors. Metallicity greatly affects the fraction of systems which move through each pathway, but generally does not change the characteristics of binaries moving through a specific pathway. From these results, we have made three observationally testable claims: (1) the majority of the ULX population can be explained as an extension of the stellar HMXB population, (2) ULXs should be preferentially found in low metallicity clusters - while the ULXs found in higher metallicity regions should be younger and have larger orbital periods, and (3) the starburst galaxy NGC 1569 is dominated by contributions from Cluster A, which likely provides approximately 90\% of the observed HMXB population. Detailed modeling of both theoretical phenomena, such as the effect of non-spherical winds falls outside the scope of this paper, but is currently ongoing. New observations of starbursts in low metallicity environments are also underway, and will have the power to quantitatively test our predictions and determine the important role that metallicity plays in HMXB evolution.

\acknowledgements
We thank Krzysztof Belczynski for many helpful comments on StarTrack models. We acknowledge support from NSF grant AST-0449558 and NASA GO0-11108B/NAS8-03060 to VK. TL also acknowledges support from the GAANN Fellowship under the Department of Education. JSG acknowledges support from NSF AST-0708967 to the University of Wisconsin-Madison. 

\bibliography{ms}

\end{document}